%% file: main.tex
\documentclass[10pt, a4paper, twocolumn, twoside]{article} 
\usepackage{graphicx}
\usepackage{subfigure}
\usepackage{amssymb}
\usepackage{float}
\usepackage{pifont}
\input{structure.tex} 

\title{The impact of pre-white dwarf evolution on the pulsational
  properties and asteroseismological inferences of ZZ Ceti
  stars} 

\shorttitle{The impact of pre-white dwarf evolution on the pulsational
  properties of ZZ Ceti
  stars} 

\shortauthors{De Ger\'onimo et al.} 

\author{
        \authorstyle{F.~C.~De Ger\'onimo,$^{1,2}$, L.~G.~Althaus,$^{1,2}$, A.~H.~C\'orsico$^{1,2}$, A.~D.~Romero$^3$ and S.~O.~Kepler$^3$}
	\newline\newline 
        $^1$\institution{Facultad de Ciencias Astron\'omicas y Geof\'isicas, Universidad
          Nacional de La Plata, Paseo del Bosque s/n, (1900) La Plata, Argentina; fdegeronimo@fcaglp.unlp.edu.ar}\\ 
	$^2$\institution{Instituto de Astrof\'isica de La Plata, CONICET-UNLP}\\ 
	$^3$\institution{Departamento de Astronomia, Universidade Federal do Rio Grande do Sul, 
          Av. Bento Goncalves 9500, Porto Alegre 91501-970, RS, Brazil; alejandra.romero@ufrgs.br}\\ 
      }

\begin{document}

\maketitle 

\thispagestyle{firstpage} 
\newabstract{ZZ Ceti stars are pulsating white dwarfs with a
  carbon-oxygen core (or possibly ONe for the most massive stars)
  build up during the core helium burning (CHeB) and thermally pulsing
  Asymptotic Giant Branch (TP-AGB) phases. Through the interpretation
  of their pulsation periods by means of asteroseismology, details
  about their origin and evolution can be inferred. The whole
  pulsation spectrum exhibited by ZZ Ceti stars strongly depend on the
  inner chemical structure. At present, there are several processes
  affecting the chemical profiles that are still not accurately
  determined. We present a study of the impact of current
  uncertainties in the evolution of white dwarf progenitor on the
  expected pulsation properties and on the stellar parameters inferred
  from asteroseismological fits of ZZ Ceti stars. Our analysis is
  based on a set of carbon-oxygen core white dwarf models that are
  derived from full evolutionary computations from the ZAMS to the ZZ
  Ceti domain. We considered models in which we varied the number of
  thermal pulses, the amount of overshooting, and the carbon-alpha
  reaction rate within their uncertainties. We explore the impact of
  these major uncertainties in prior evolution on the chemical
  structure and expected pulsation spectrum. We find that these
  uncertainties yield significant changes in the g-mode pulsation
  periods being those found during the TP-AGB phase the most relevant
  for the pulsational properties and the asteroseismological derived
  stellar parameters of ZZ Ceti stars.  We conclude that the
  uncertainties in the white dwarf progenitor evolution should be
  taken into account in detailed asteroseismological analyses of these
  pulsating stars.
 
  }


\section{Introduction}
ZZ Ceti (or DAV) stars are pulsating white dwarfs (WDs) with H-rich
atmospheres. Located in a narrow instability strip with effective
temperatures between 10500 K and 12500 K
\citep{2008PASP..120.1043F,2008ARA&A..46..157W,2010A&ARv..18..471A},
they constitute the most numerous class of compact pulsators. Their
luminosity variations are due to non-radial {\it g}-mode pulsations of
low degree ($\ell \leq 2$) with periods between 70 and 1500 s and
amplitudes up to 0.30 mag, excited by both the $\kappa-\gamma$
mechanism \citep{1981A&A...102..375D,1982ApJ...252L..65W} and the
``convective driving'' mechanism
\citep{1991MNRAS.251..673B,1999ApJ...511..904G,2013EPJWC..4305005S}.

Through asteroseismology, we are able to infer details about the
progenitor evolution
\citep{2008PASP..120.1043F,2008ARA&A..46..157W,2010A&ARv..18..471A}.
Nowadays, two main approaches are adopted for WD asteroseismology. The
first one employs static stellar models with parametrized chemical
profiles
\citep{2011ApJ...742L..16B,2014ApJ...794...39B,2014IAUS..301..285G,2016ApJS..223...10G}.
The second one, our asteroseismological approach, adopt fully
evolutionary models that result to the complete evolution of the
progenitor star, from the Zero Age Main Sequence (ZAMS) to the WD
stage \citep{2012MNRAS.420.1462R,2013ApJ...779...58R}, but inferences
based on this approach do not take into account uncertainties neither
in the modeling nor in the input physics of WD progenitors. Both affect
the shape of the chemical abundance distribution which is critical for
the pulsational properties of WDs.  In particular, the core helium
burning phase is affected by two main uncertainties: the
$^{12}$C$(\alpha,\gamma)^{16}$O reaction rate and the amount of
overshooting (OV) adopted.
In addition, the chemical structure of the outermost part of the core
depend on the occurrence of the thermal pulses, those who in turn depend on
 both the amount of OV adopted during the TP-AGB phase
and the poorly constrained efficiency of mass loss
\citep{2014PASA...31...30K}.




\begin{figure*}[ht]
  \centering
  \mbox{
    \subfigure[\label{pt-perfil-0548}]{\includegraphics[width=6cm, angle=270]{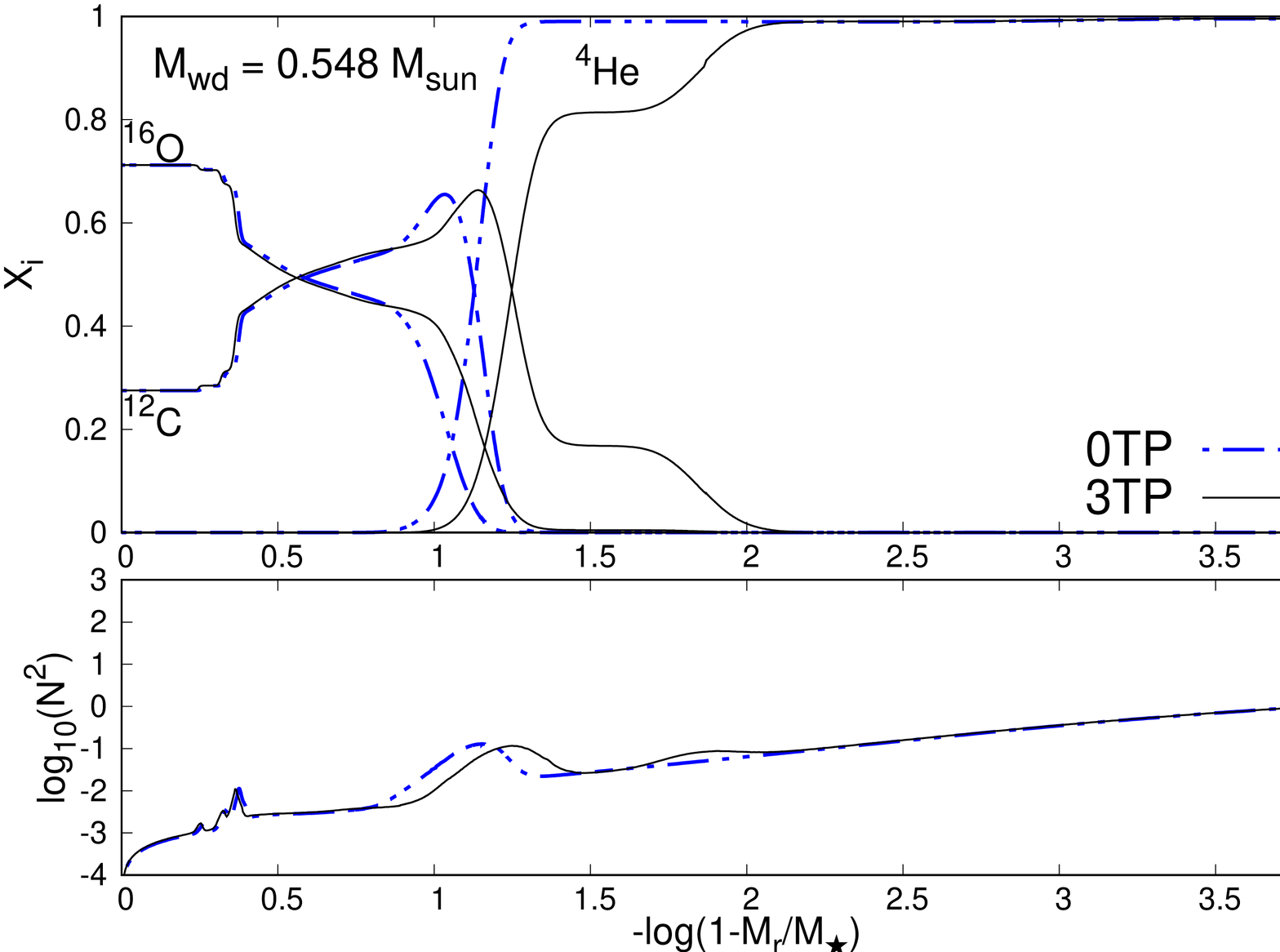}}\quad
    \subfigure[\label{pt-perfil-0837}]{\includegraphics[width=6cm, angle=270]{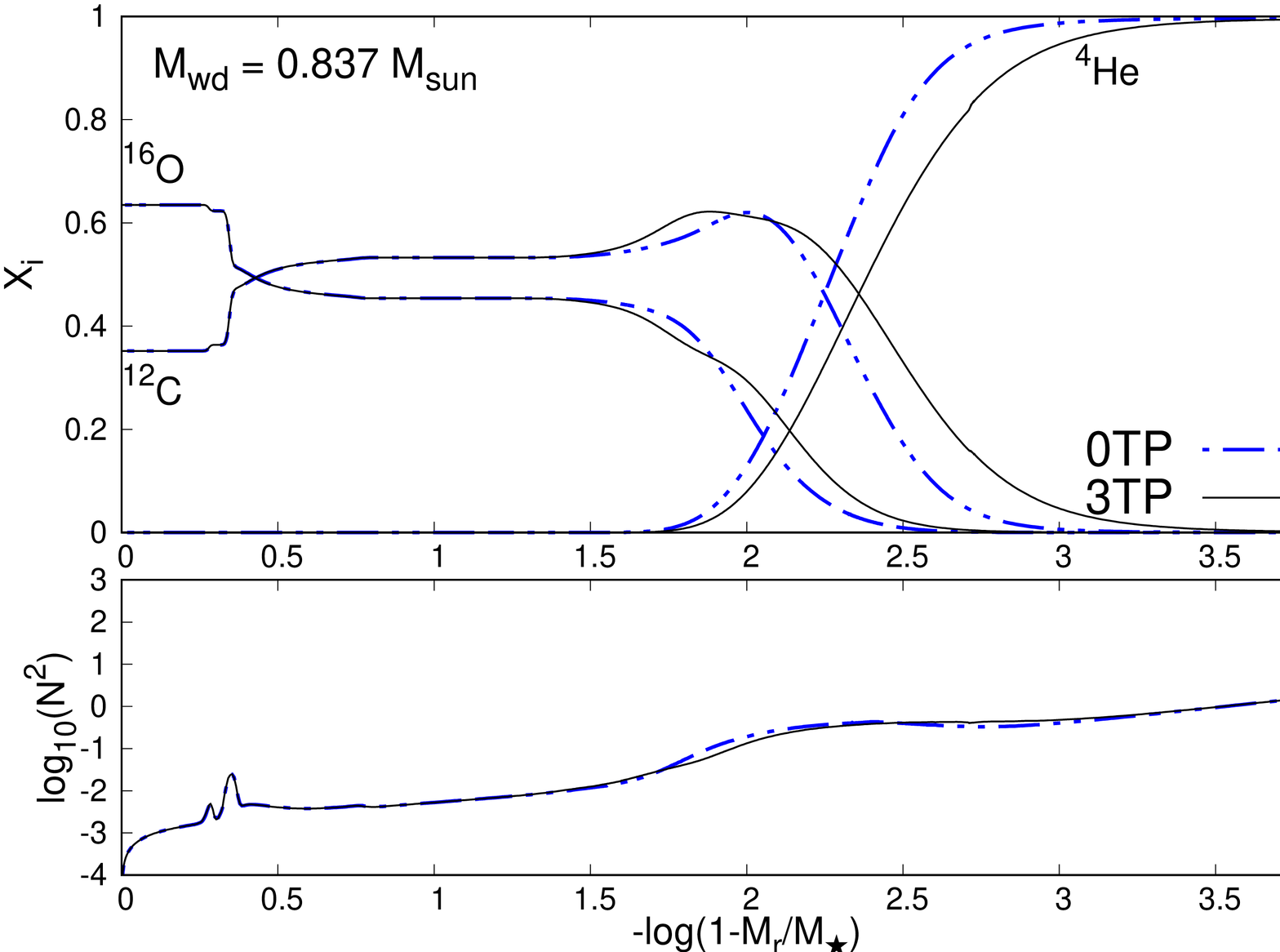}}\quad
    
  }
  \caption{Inner O, C and He abundance distribution in
    terms of the outer mass fraction for the two stellar models 
    considered at $T_{\rm eff}\sim 12\,000$ K (upper panels) and the Brunt-V\"ais\"al\"a frequency (lower panels). 
  }
  \label{pt-perfil}
\end{figure*}

    

\begin{figure*}[ht]
  \centering
  \mbox{
    \subfigure[\label{pt-Pdif-0548}]{\includegraphics[width=6cm, angle=270]{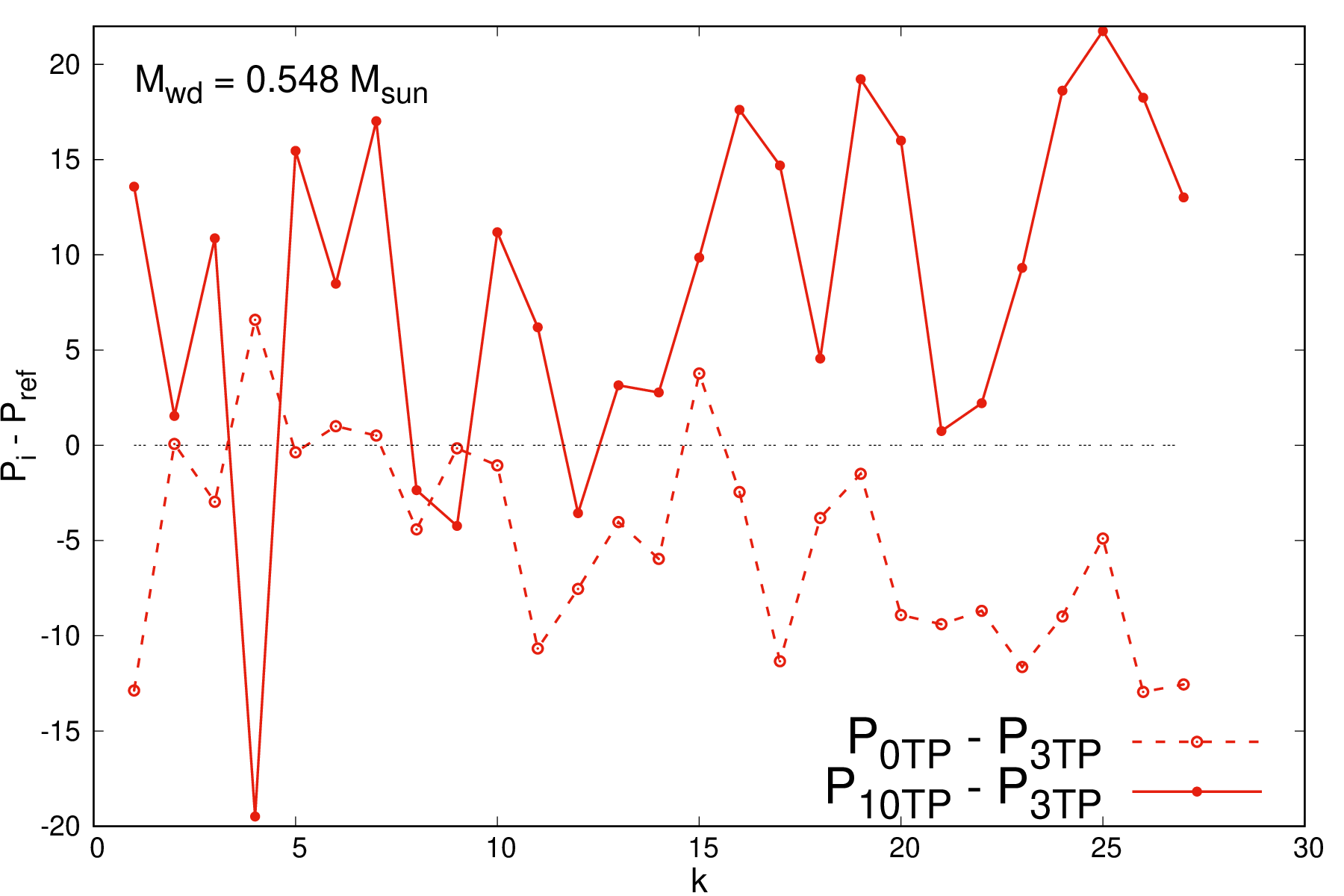}}\quad
    \subfigure[\label{pt-Pdif-0837}]{\includegraphics[width=6cm, angle=270]{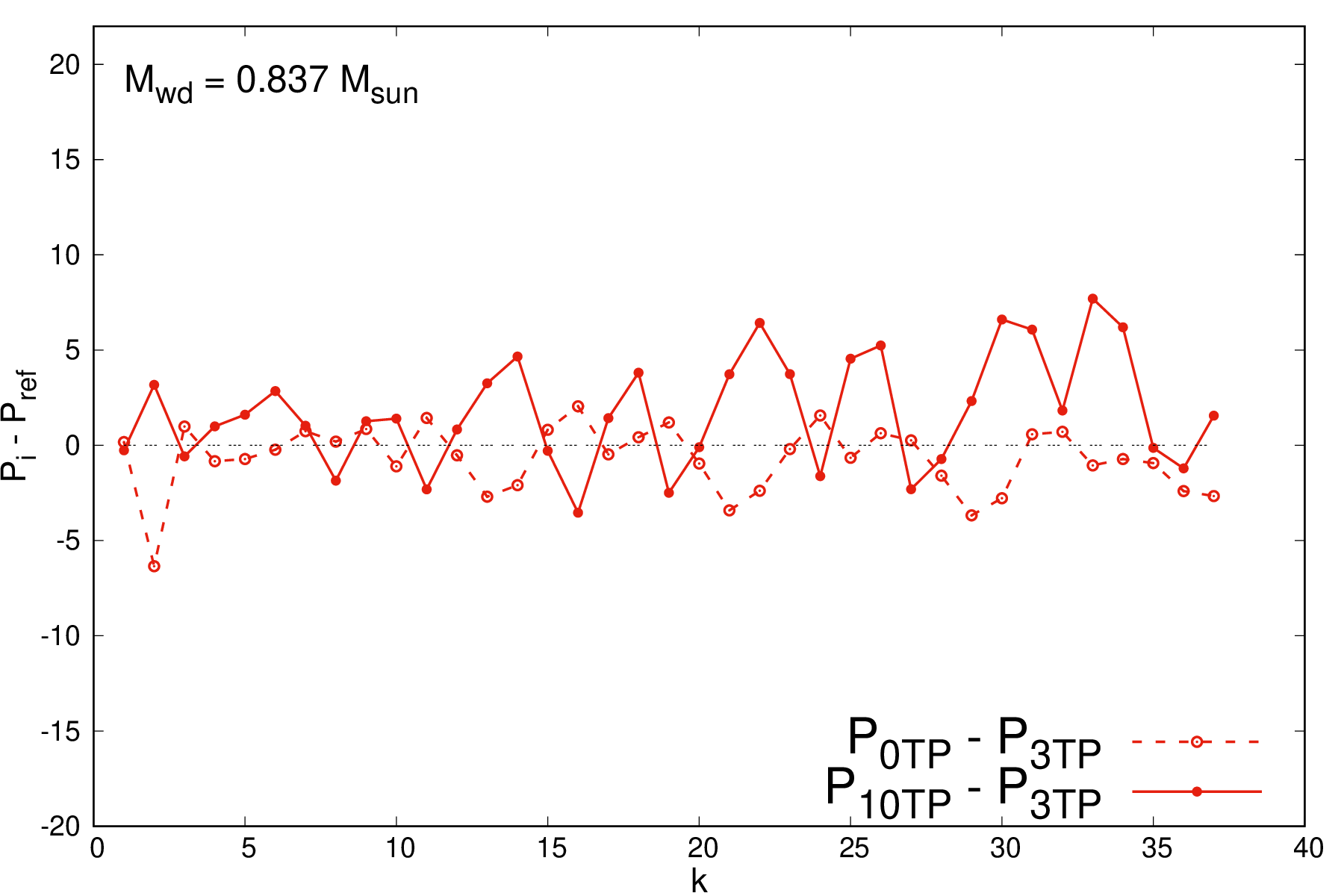}}\quad
    
  }
  \caption{Period differences in terms of $k$ between the 0TP and 3TP
    models and between the 10TP and 3TP models for the
    $M_{\star}= 0.548 M_{\odot}$ and $M_{\star}= 0.837 M_{\odot}$ WD
    models (left and right panels, respectively).}
  \label{pt-Pdif}
\end{figure*}

\begin{figure*}[ht]
\centering
  \mbox{
    \subfigure[\label{ov-perfil-0548}]{\includegraphics[width=6cm, angle=270]{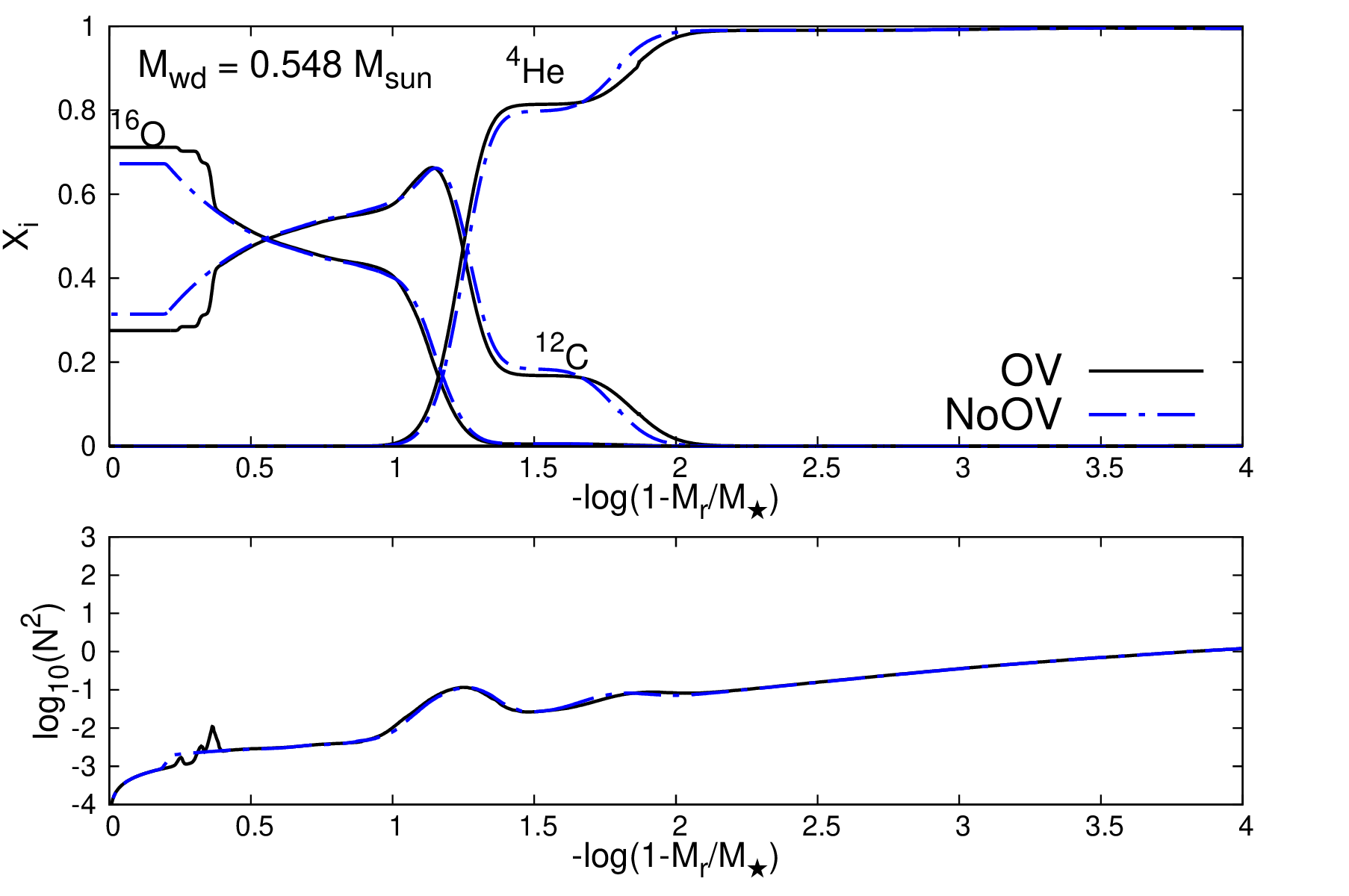}}\quad
    \subfigure[\label{ov-Pdif-0548}]{\includegraphics[width=6cm, angle=270]{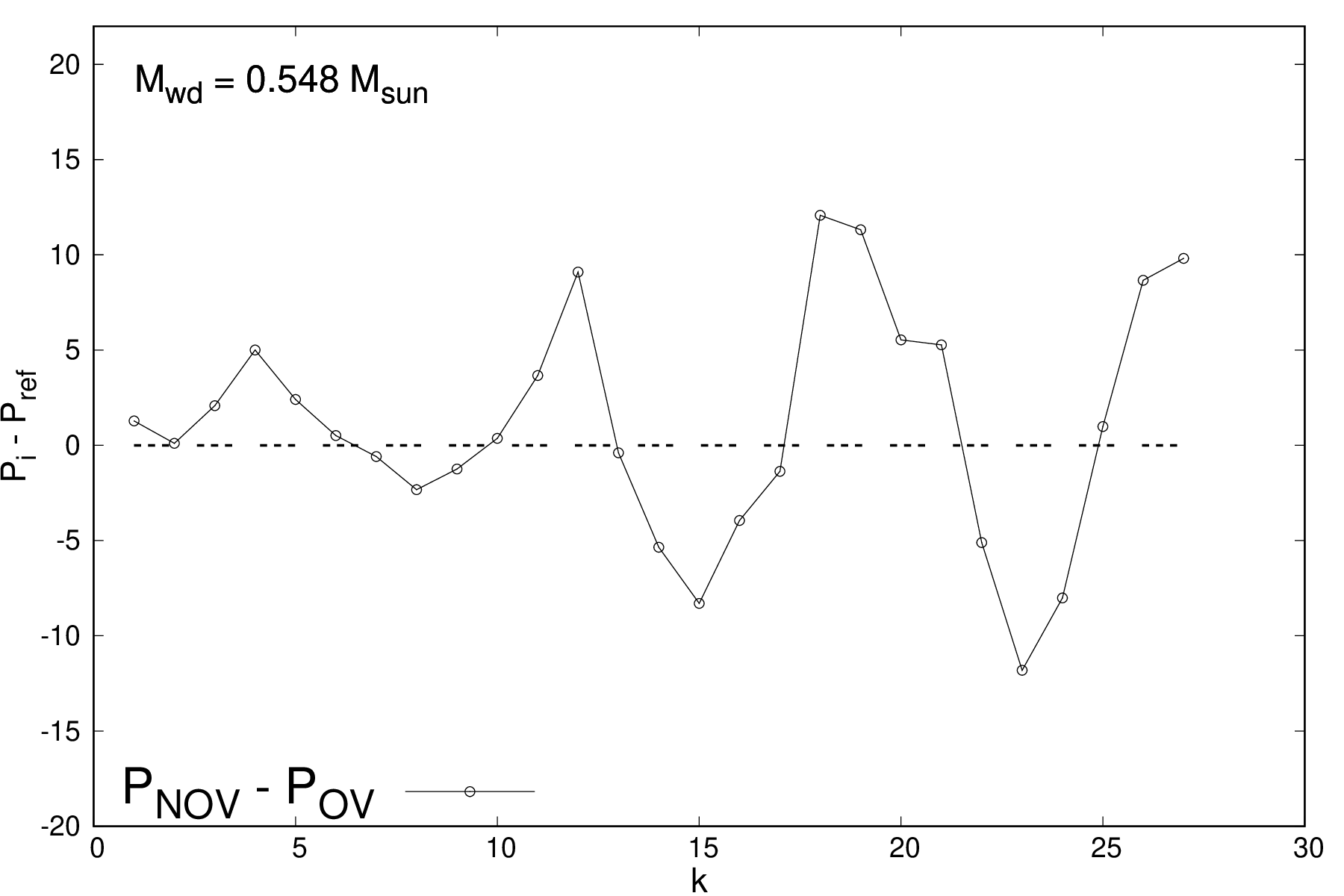}}\quad
    
  }
  \caption{Inner chemical profiles for models computed with an OV
    parameter $f= 0.016$ (OV) and with $f= 0$ (NOV) during core He
    burning (left panel) and the period differences for $k$ fixed, for
    the low mass model.}
  \label{ov-perfil}
\end{figure*}

    

\begin{figure*}
  \centering \mbox{
    \subfigure[\label{co-perfil-0548}]{\includegraphics[width=6cm,
      angle=270]{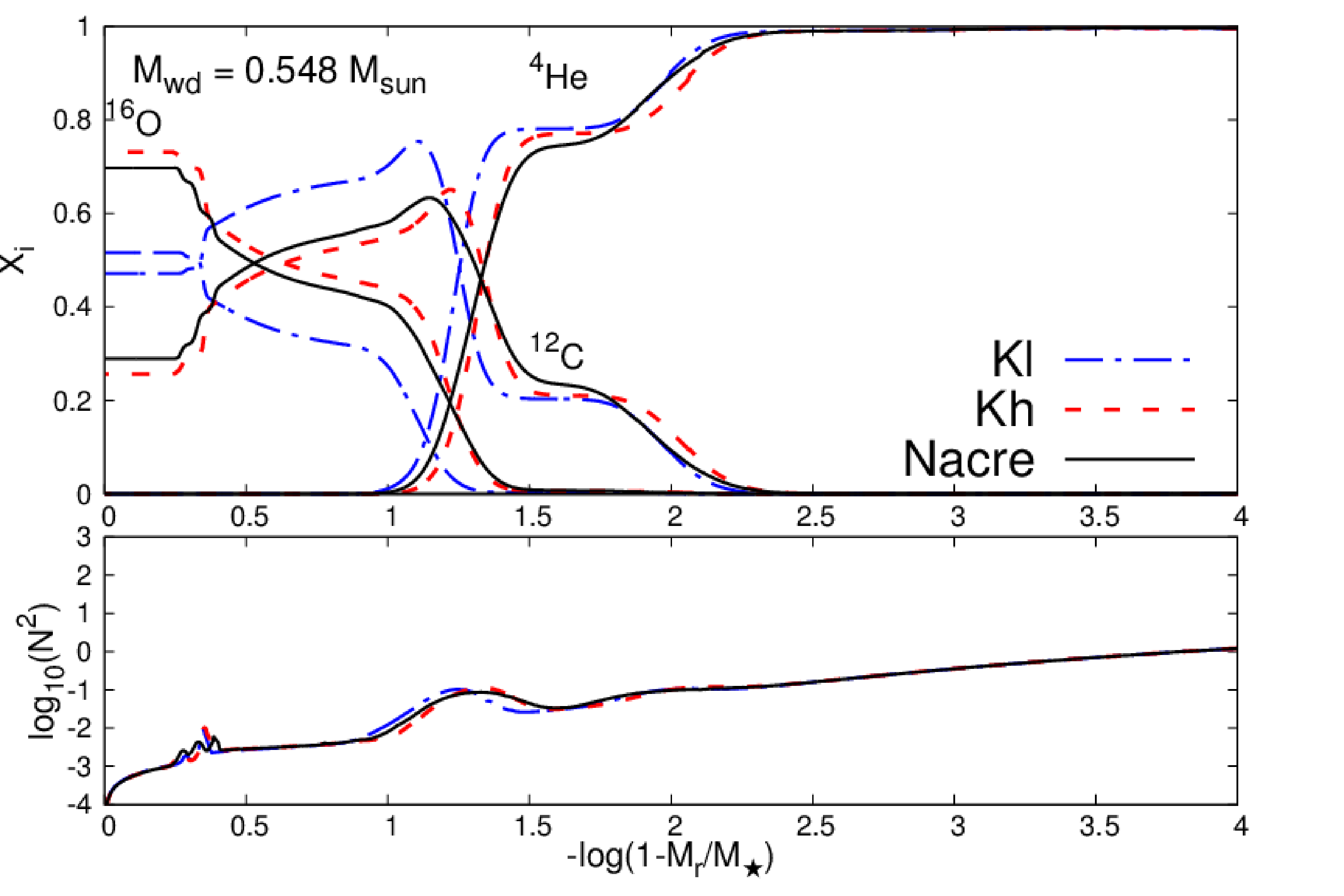}}\quad
    \subfigure[\label{co-Pdif-0548}]{\includegraphics[width=6cm,
        angle=270]{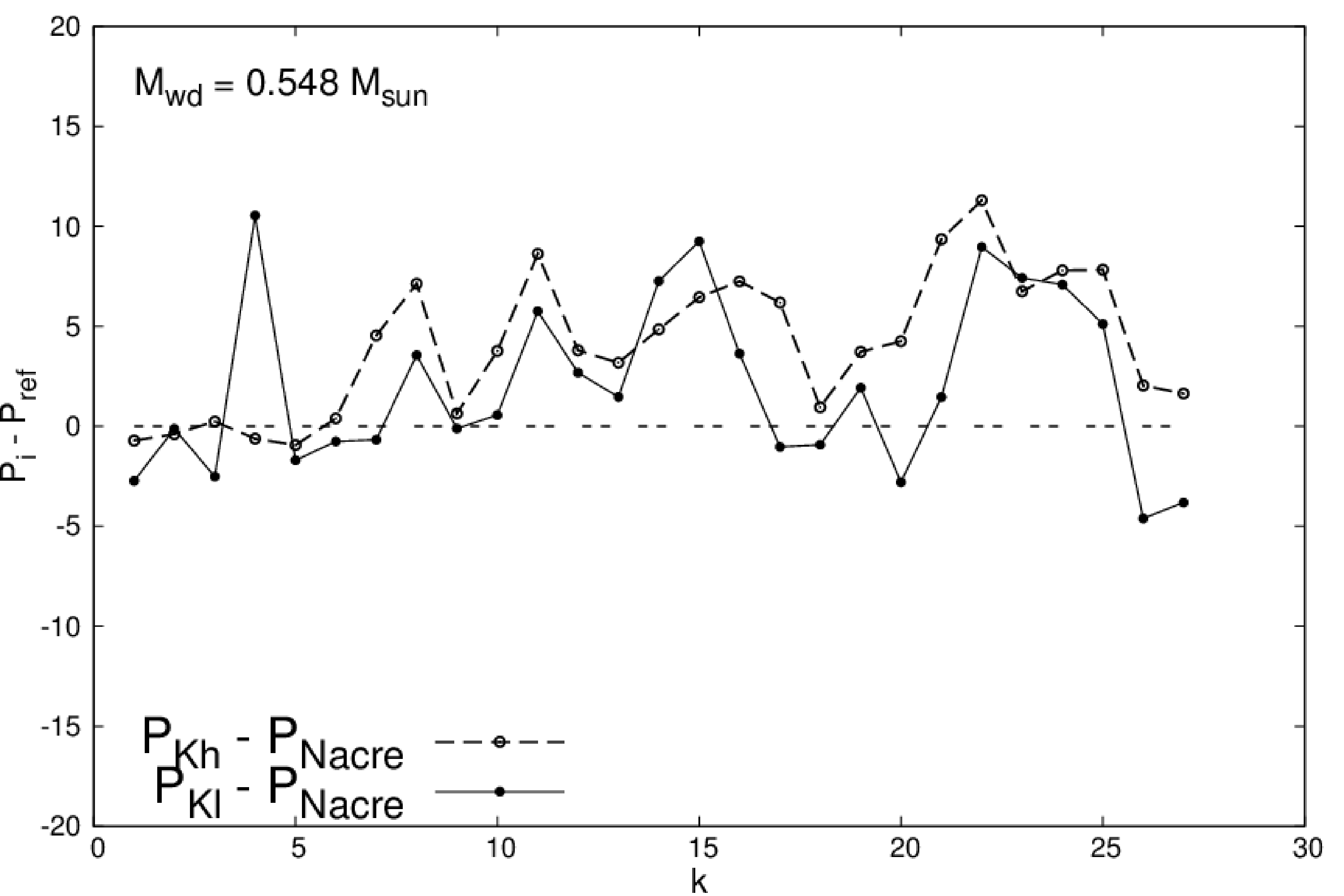}}\quad
  }
  \caption{Same as Fig. \ref{ov-perfil} but for models computed with
    three different $^{12}$C$+\alpha$ reaction rates during CHeB (left panel) and
    the corresponding period differences (right panel).  }
  \label{co-perfil}
\end{figure*}
    

\section{Impact over the chemical structure and period spectrum}

As a first step, we explore the impact of uncertainties linked to:

\begin{enumerate}
\item the occurrence of TP during the AGB phase of the
progenitor star, \label{it:TP}
\item the occurrence of OV during core He burning, \label{it:OV}
\item the $^{12}$C$(\alpha,\gamma)^{16}$O reaction rate, \label{it:CO}
\end{enumerate}

\noindent
on the chemical structure and period spectra of template models
characterized by $M_{\star}= 0.548, 0.837 M_{\odot}$, with H envelope
mass of $M_{\rm H} \sim 4\times 10^{-6} M_{\star}$ at $T_{\rm eff}
\sim 12\,000$ K, and considered the pulsation period spectrum for
modes with $\ell= 1$. All our evolutionary computations were performed
with the LPCODE evolutionary code, while the pulsational properties of
our models were computed with LP-PUL, both numerical tools developed in
La Plata, Argentina.



\subsection{Occurrence of TP in AGB}

Because during this phase a pulse-driven convection zone is developed, the
building of an intershell region and a double-layered chemical
structure is expected at the bottom of the He buffer. The size of this
region decreases as the star evolves
through the TP-AGB, as consequence of the He-burning shell. The number
of TP experienced by the progenitor star is uncertain and depends on
the rate at which mass is lost during the TP-AGB phase, on the initial
metallicity of progenitor star, as well as on the occurrence of
extra-mixing in the pulse-driven convection zone
\citep{2000A&A...360..952H}.
We performed a comparison between the chemical structure and period
spectra of models at the ZZ Ceti stage, whose progenitors depart from
the AGB previous to the first thermal pulse (0TP model) and at the end
of the third and tenth thermal pulse (referenced as 3TP and 10 TP
models).
Fig. \ref{pt-perfil} display the chemical profiles (upper panel)
for the most abundant elements, and the logarithm of the squared 
Brunt-V\"ais\"al\"a
(B-V) frequency (lower panel) in terms of the outer mass. 
Diffusion processes are more efficient for high-mass model, eroding
the intershell region by the time that the model reaches the ZZ Ceti
stage resulting in a lower impact on the period spectra, shown in Fig
\ref{pt-Pdif}.
There we show the impact of considering different number of TP on the
period spectrum of the models, that is, the period differences
($\Delta P_{k} \equiv P_{k, {\rm 0TP}} -P_{k, {\rm 3TP}}$) between the
0TP and 3TP models and the period differences
($\Delta P_k \equiv P_{k, {\rm 10TP}} -P_{k, {\rm 3TP}} $) between the
10TP and 3TP models as a function of the radial order $k$.  These
substancial variations in the less massive model, with average
differences of 10 s, are not solely due to presence of the intershell
region (or the overall shape of the profile), but also due to the
outward shift of the core/He main chemical transition region.
In the other hand, the expectations for the massive model are markedly
different. Indeed, in this case the period differences are on average
3 s, resulting to the slight differences found in chemical profiles.
In fact, no intershell is expected to occur at the ZZ Ceti domain.


\subsection{Occurrence of OV during core He burning}

Extra mixing episodes during core He burning strongly affect the final
shape of the carbon-oxygen core.  We performed a comparison between
models in the ZZ Ceti stage, that comes from progenitor stars in which
we considered the extreme situation of no OV allowed during core He
burning (NOV model), and compared it to the one in which we considered
a moderate OV (OV model).  Fig. \ref{ov-perfil-0548} show the inner
chemical profiles at the ZZ Ceti stage. The ingestion of fresh He in a
C-rich zone, due to the inclusion of OV, favour the consumption of C
via the reaction $^{12}$C$+\alpha$, with the consequent increase of
$^{16}$O.  The B-V frequency is notoriously affected by the steep
variation in the carbon-oxygen profile left by core OV.  Differences
in the period spectra are shown in Fig. \ref{ov-Pdif-0548}, revealing
variations of 4 s on average, markedly lower than the period
differences resulting from uncertainties in the TP-AGB phase, being
the period differences resulting from the inclusion of OV, similar for
both stellar masses.

\subsection{Uncertainties in the $^{12}$C$(\alpha,\gamma)^{16}$O reaction
  rate}

In this case, we computed the CHeB phase of the progenitors
by considering three different values for the
$^{12}$C$(\alpha,\gamma)^{16}$O reaction rate, and followed the
evolution until the progenitor leave the TP-AGB phase.  We considered
the reaction rates provided by \citet{1999NuPhA.656....3A} (Nacre, our
reference model), and the extreme high and low values from
\citet{2002ApJ...567..643K} (Kh and Kl models, respectively). These
values for the reaction rate allow us to account for the
uncertainties in the carbon-oxygen chemical profiles of WD. For core
He burning temperatures $\sigma_{\rm Kl}/ \sigma_{\rm Nacre}\sim 0.55$
and $\sigma_{\rm Kh}/ \sigma_{\rm Nacre}\sim 1.1$, being the
uncertainty factor of about 2.  Fig. \ref{co-perfil-0548} shows the
chemical profiles for the models at the ZZ Ceti stage whose
progenitors were computed by adopting different reaction rates during
CHeB.  Differences in chemical structure are mainly found in the
location of the O/C and O/C/He chemical transitions and in the
abundances of the intershell region, with the resulting change in the
B-V frequency.

Fig. \ref{co-Pdif-0548} shows the differences in the pulsation spectrum due
to the uncertainty in the $^{12}C+\alpha$ reaction rate, by taking as
reference the model computed with the Nacre value. In particular, for
the low-mass model, appreciable differences are expected for both low
and high radial-order modes, with average variations of 4 s, while in
the case of the massive model (not showed here), small period
differences are found for modes with radial orders $k < 23$. These
small variations reflect the smooth behaviour of the chemical profile
at the outermost part of the carbon-oxygen core.  But, for radial
orders $k > 23$, appreciable differences are found, that stem from
the fact that some modes of high radial order $k$ are sensitive to the
chemical structure of the core.

\section{Impact over asteroseismological determinations}



As we have seen before,
, whether the progenitor star evolves through the
TP-AGB or not, strongly impact the period spectrum of pulsating DA
WDs, specially in the case of the low-mass ZZ Ceti stars.

For our study of the impact of the thermal pulses on the stellar
parameters derived through asteroseismology, we developed a grid of
evolutionary models with final masses in the range
$0.5349 \lesssim M_{\rm wd}/M_{\odot} \lesssim 0.6463$.
During the progenitor evolution we forced the evolutionary models to
abandon this stage by enhancing the stellar mass loss rate at two
different stages: previous to the onset of the first thermal pulse,
and at the end of the third thermal pulse (0TP and 3TP models,
respectively). 
We want to mention that, it is during the first TP that the main
chemical features of the intershell and the double layered regions
emerge. In the case of low mass stars, we do not expect the occurrence
of a large number of TP, so considering 3 TP is enough for capturing
the essential of the chemical structure arising from TP-AGB
phase. 
For higher stellar masses, we expect a larger number of TPs to take
place, but as shown before, see also \citet{2017A&A...599A..21D}, the
period spectrum expected at the ZZ Ceti stage is not markedly affected
by the number of additional TPs experienced by the progenitor
star.

To study the impact of the uncertainties in
the $^{12}C+\alpha$ reaction rate on the asteroseismological-derived
stellar parameters, we considered a stellar grid of models with masses
$0.523 \lesssim M_{\rm wd}/M_{\odot} \lesssim 0.90$, for which we take
into account, during CHeB phase, both low- and high-values of the rate
for the $^{12}C+\alpha$ nuclear reaction from the work of
\citet{2002ApJ...567..643K} obtaning two sets of models ($K_l$ and
$K_h$, respectively).

In addition, we considered H-envelopes thinner than those predicted
from evolutionary computations, ranging
$-9\lesssim log (M_{\rm H}/M_{\rm wd}) \lesssim -4$.

\begin{equation}
\phi=\phi(M_{\star},M_{\rm H},T_{\rm eff})= \frac{1}{N}\sqrt{\sum_{i=1}^N\frac{[\Pi_i^{th}-\Pi_i^{obs}]^2A_i}{\sum_{i=1}^N A_i}}
\label{eq:phi}
\end{equation}
As a first approach, we performed period-to-period fits to 1000
artificially-generated pulsating DA WD\footnote{Our
  artificially-generated pulsating DA are composed by 3-random
  generated periods, with values representative of ZZ Ceti
  stars.}. That is, we selected the best fit model for each set of
models, 0TP and 3TP (or $K_l$ and $K_h$) according to the minima of
the quality function $\phi$ [Eq. (\ref{eq:phi})] --- the best
period-fit model--.  
By making histograms with the value of the differences in the
asteroseismologically-derived stellar parameters and adjusting to them
Gaussian function we found that the sigma ($\sigma$) values are very
close to the observational errors for those quantities
\citep{2011ApJ...730..128T}, see table \ref{tabla:diferencias} \citep[for more
details, see][]{2018A&A...613A..46D}.

\begin{table}[]
\resizebox{\columnwidth}{!}{
\begin{tabular}{lll}
  \hline
{\color{blue}$\sigma_{Teff}$}\\
\hline
  $T_{eff}$    & Thermal pulses &$^{12}C+\alpha$ \\
  \hline
  hot          &    5.8\%         &    5\%            \\
  intermediate &    4.5\%         &    6\%             \\
  cool         &    8\%         &    8\%           \\
  \hline
{\color{blue}$\sigma_{Mwd}$}\\

\hline 
 hot          &    7\%       &        3\%         \\
  intermediate &   6\%          &      2\%         \\
  cool         &    11\%         &     20\%          \\
  \hline
{\color{blue}$\sigma_{Mh}$} (order of magnitude)\\

\hline 
hot          & 0.5            &     0.6            \\
intermediate & 0.6            &     0.74            \\
cool         &  2          &     2           \\
\hline

\end{tabular}}
\caption{$\sigma$ values for the deviation in the stellar parameters
  for hot-, intermediate- cool-$T_{eff}$ artifficially-generated ZZ
  Ceti. Values are strongly improved if $M_H$ remain fixed}
\label{tabla:diferencias}
\end{table}

Finally, we contrast the previous results with those derived from
asteroseismological fits to some selected real ZZ Ceti stars. We
considered those ZZ Ceti stars with modes previously identified as
$\ell=1$, with spectroscopic masses close to the mass range
considered in this work. 
We classify these stars as cool, intermediate-$T_{\rm eff}$ or hot ZZ
Ceti stars depending on the value of the pulsation period with the
highest amplitude.  For the selection of the best fit models we have
taken into account:

\begin{itemize}
\item  the models minimize the quality function given by Eq. \ref{eq:phi}, 
\item  $T_{\rm eff}$ and $\log(g)$,
as close as possible to the spectroscopic determinations.
\end{itemize}

\begin{figure*}[h]
\centering
  \includegraphics[width=1\textwidth]{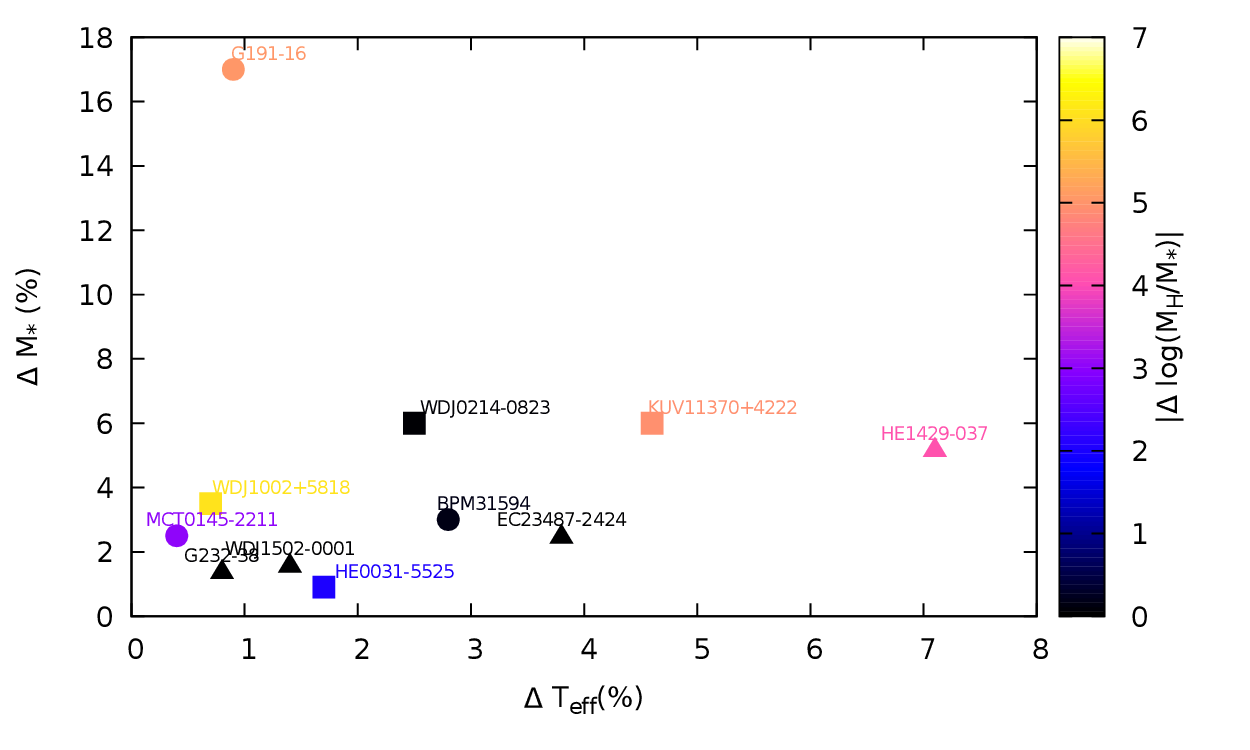}
  \caption{Variations of $T_{\rm eff}$, $M_{\rm wd}$ and $M_{\rm H}$
    --in color scale-- of selected ZZ Ceti stars resulting from the
    two sets of evolutionary sequences correspondig to the study of
    thermal pulses. Squares, circles, and triangles refer to hot,
    intermediate, and cool ZZ Ceti stars, respectively.}
\label{pt-real-zzceti}
\end{figure*}

\begin{figure*}[h]
\centering
\includegraphics[width=1\textwidth]{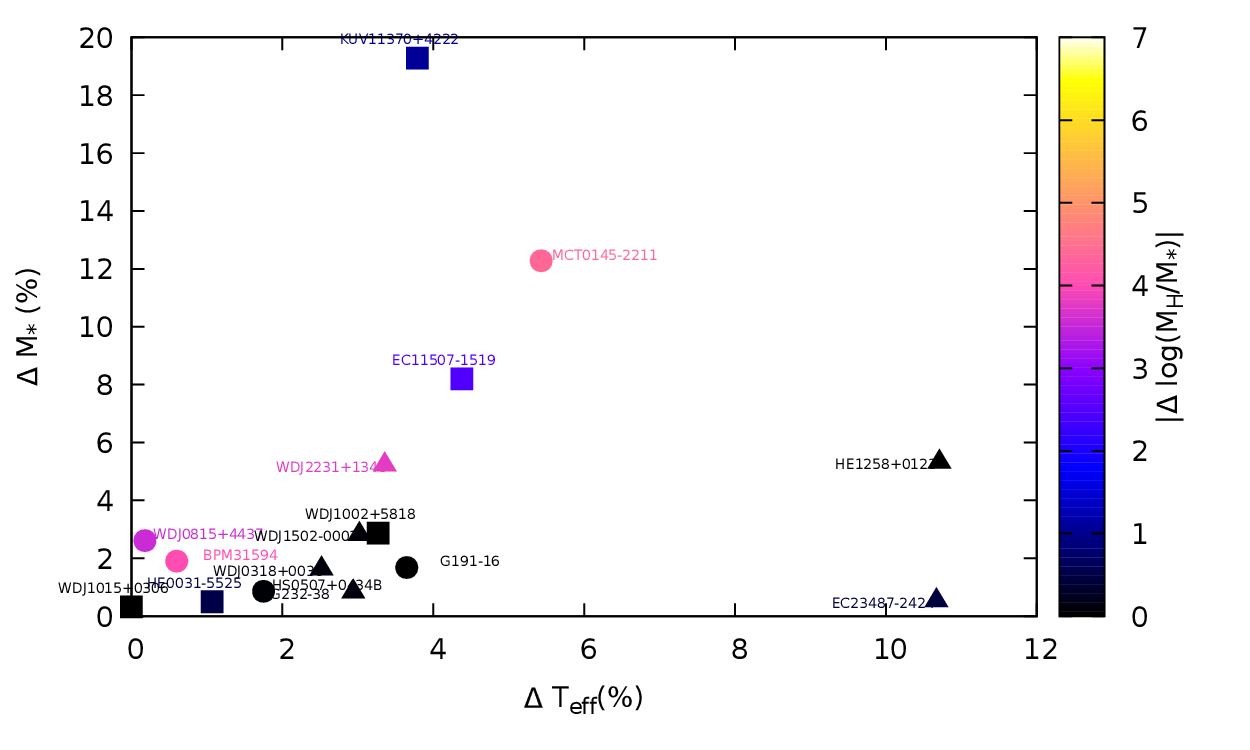}
\caption{Same as Fig.\ref{pt-real-zzceti} but for the study of the $^{12}C(\alpha,\gamma)^{16}O$ reaction rate.}
\label{co-real-zzceti}
\end{figure*}

The results of the asteroseismological period-to-period fits to the
real ZZ Ceti stars are illustrated in Figs. \ref{pt-real-zzceti} and
\ref{co-real-zzceti}. The figures show the deviations found in the
asteroseismologically-derived stellar parameters as given by the two
sets of evolutionary sequences. In the X-axis we plot the percentage
of the deviation in effective temperature, in the Y-axis the deviation
in stellar mass while in color scale the absolute difference in
$log(M_H/M_{wd} )$. From the figures we warn that these results are
consistent with those found
previoulsy. 
 Although we might expect a certain link between the magnitude of the
deviations found on the different stellar parameters, we did not found
a clear pattern.  An important result is that some hot- and
intermediate-effective temperature ZZ Ceti stars are sensitive to the
thickness of the hydrogen envelope.

\section{Conclusions}

Marked differences in the pulsational period spectrum are expected due
to the uncertainties in the $^{12}C(\alpha,\gamma)^{16}O$ reaction
rate and whether the progenitor star evolves or not through the TP-AGB
phase. We find that these uncertainties during the progenitor
evolution have a non-negligible impact on the resulting
asteroseismologically-derived stellar parameters. We report that these
uncertainties implies average deviations close to the spectroscopic
errors, except for cool ZZ Ceti where the uncertainties are
appreciably greater. In the other hand, for the mass of the H envelope
we find deviations up to 2 order of magnitude in the case of
low-$T_{eff}$ ZZ Ceti models, while for hot- and
intermediate-$T_{eff}$ ZZ Ceti models, we find negligible differences.
We want to highlight that these results are obtained by considering
extreme situations, both for the values of the reaction rate as well
as the evolution during the TP-AGB phase. We also find that the period
spectrum of some hot and intermediate effective temperature ZZ Ceti
stars are sensitive to the mass of the hydrogen envelope.  Our main
finding is that the uncertainties in prior WD evolution do affect both
the period spectra of pulsating DA WD and WD asteroseismology, but the
effects are quantifiable and bounded. In particular, differences found
in the determinations of stellar mass, effective temperature and H
envelope thickness due to the uncertainties studied, are within the
typical spectroscopic errors. These results add confidence to the use
of fully evolutionary models with consistent chemical profiles, and
render much more robust our asteroseismological approach.


\bibliography{papers}

\end{document}

%% file: structure.tex
\usepackage[english]{babel} 

\usepackage{microtype} 

\usepackage{amsmath,amsfonts,amsthm} 

\usepackage[svgnames]{xcolor} 

\usepackage[small, labelfont=bf, up]{caption} 

\usepackage{booktabs} 

\usepackage{lastpage} 

\usepackage{graphicx} 

\usepackage{enumitem} 
\setlist{noitemsep} 

\usepackage{sectsty} 
\allsectionsfont{\usefont{OT1}{phv}{b}{n}} 

\usepackage{geometry} 

\usepackage[T1]{fontenc} 
\usepackage[utf8]{inputenc} 

\usepackage{XCharter} 

\usepackage{hyperref}

\usepackage{journals}

\usepackage{fancyhdr} 
\newcommand{\shorttitle}[1]{\fancyhead[CE]{\textsl{#1}}}
\newcommand{\shortauthors}[1]{\fancyhead[CO]{\textsl{#1}}}

\fancypagestyle{firstpage}{ 
	\fancyhf{}
        \lhead{\vspace*{0.0em} \textit{\footnotesize 21$^{\rm st}$ European Workshop on White
            Dwarfs \\ Castanheira, Campos, Montgomery, eds. \\[-0.2em]
          July 23--27, 2018, Austin, Texas, USA}}
}

\date{}

\newcommand{\authorstyle}[1]{{\large\usefont{OT1}{phv}{b}{n}\color{DarkRed}#1}} 

\newcommand{\institution}[1]{{\footnotesize\usefont{OT1}{phv}{m}{sl}\color{Black}#1}} 

\usepackage{titling} 

\newcommand{\HorRule}{\color{DarkGoldenrod}\rule{\linewidth}{1pt}} 

\pretitle{
	\vspace{-30pt} 
	\HorRule\vspace{10pt} 
	\fontsize{16}{12}\usefont{OT1}{phv}{b}{n}\selectfont 
	\color{DarkRed} 
}

\posttitle{\par\vskip 10pt} 

\preauthor{} 

\postauthor{ 
	\vspace{0pt} 
	{\par\HorRule} 
	\vspace{-10pt} 
}

\newcommand{\newabstract}[1]{
    {\section*{Abstract}
    \bfseries #1}
  }

\usepackage[authoryear]{natbib}

%% file: main.bbl
\begin{thebibliography}{}
\makeatletter
\relax
\def\mn@urlcharsother{\let\do\@makeother \do\$\do\&\do\#\do\^\do\_\do\%\do\~}
\def\mn@doi{\begingroup\mn@urlcharsother \@ifnextchar [ {\mn@doi@}
  {\mn@doi@[]}}
\def\mn@doi@[#1]#2{\def\@tempa{#1}\ifx\@tempa\@empty \href
  {http://dx.doi.org/#2} {doi:#2}\else \href {http://dx.doi.org/#2} {#1}\fi
  \endgroup}
\def\mn@eprint#1#2{\mn@eprint@#1:#2::\@nil}
\def\mn@eprint@arXiv#1{\href {http://arxiv.org/abs/#1} {{\tt arXiv:#1}}}
\def\mn@eprint@dblp#1{\href {http://dblp.uni-trier.de/rec/bibtex/#1.xml}
  {dblp:#1}}
\def\mn@eprint@#1:#2:#3:#4\@nil{\def\@tempa {#1}\def\@tempb {#2}\def\@tempc
  {#3}\ifx \@tempc \@empty \let \@tempc \@tempb \let \@tempb \@tempa \fi \ifx
  \@tempb \@empty \def\@tempb {arXiv}\fi \@ifundefined
  {mn@eprint@\@tempb}{\@tempb:\@tempc}{\expandafter \expandafter \csname
  mn@eprint@\@tempb\endcsname \expandafter{\@tempc}}}

\bibitem[\protect\citeauthoryear{{Althaus}, {C{\'o}rsico}, {Isern}  \&
  {Garc{\'{\i}}a-Berro}}{{Althaus} et~al.}{2010}]{2010A&ARv..18..471A}
{Althaus} L.~G.,  {C{\'o}rsico} A.~H.,  {Isern} J.,   {Garc{\'{\i}}a-Berro} E.,
   2010, \mn@doi [\aapr] {10.1007/s00159-010-0033-1}, \href
  {http://adsabs.harvard.edu/abs/2010A%26ARv..18..471A} {18, 471}

\bibitem[\protect\citeauthoryear{{Angulo} et~al.,}{{Angulo}
  et~al.}{1999}]{1999NuPhA.656....3A}
{Angulo} C.,  et~al., 1999, \mn@doi [Nuclear Physics A]
  {10.1016/S0375-9474(99)00030-5}, \href
  {http://adsabs.harvard.edu/abs/1999NuPhA.656....3A} {656, 3}

\bibitem[\protect\citeauthoryear{{Bischoff-Kim} \&
  {{\O}stensen}}{{Bischoff-Kim} \& {{\O}stensen}}{2011}]{2011ApJ...742L..16B}
{Bischoff-Kim} A.,  {{\O}stensen} R.~H.,  2011, \mn@doi [\apjl]
  {10.1088/2041-8205/742/1/L16}, \href
  {http://adsabs.harvard.edu/abs/2011ApJ...742L..16B} {742, L16}

\bibitem[\protect\citeauthoryear{{Bischoff-Kim}, {{\O}stensen}, {Hermes}  \&
  {Provencal}}{{Bischoff-Kim} et~al.}{2014}]{2014ApJ...794...39B}
{Bischoff-Kim} A.,  {{\O}stensen} R.~H.,  {Hermes} J.~J.,   {Provencal} J.~L.,
  2014, \mn@doi [\apj] {10.1088/0004-637X/794/1/39}, \href
  {http://adsabs.harvard.edu/abs/2014ApJ...794...39B} {794, 39}

\bibitem[\protect\citeauthoryear{{Brickhill}}{{Brickhill}}{1991}]{1991MNRAS.251..673B}
{Brickhill} A.~J.,  1991, \mn@doi [\mnras] {10.1093/mnras/251.4.673}, \href
  {http://adsabs.harvard.edu/abs/1991MNRAS.251..673B} {251, 673}

\bibitem[\protect\citeauthoryear{{De Ger{\'o}nimo}, {Althaus}, {C{\'o}rsico},
  {Romero}  \& {Kepler}}{{De Ger{\'o}nimo} et~al.}{2017}]{2017A&A...599A..21D}
{De Ger{\'o}nimo} F.~C.,  {Althaus} L.~G.,  {C{\'o}rsico} A.~H.,  {Romero}
  A.~D.,   {Kepler} S.~O.,  2017, \mn@doi [\aap] {10.1051/0004-6361/201629806},
  \href {http://adsabs.harvard.edu/abs/2017A%26A...599A..21D} {599, A21}

\bibitem[\protect\citeauthoryear{{De Ger{\'o}nimo}, {Althaus}, {C{\'o}rsico},
  {Romero}  \& {Kepler}}{{De Ger{\'o}nimo} et~al.}{2018}]{2018A&A...613A..46D}
{De Ger{\'o}nimo} F.~C.,  {Althaus} L.~G.,  {C{\'o}rsico} A.~H.,  {Romero}
  A.~D.,   {Kepler} S.~O.,  2018, \mn@doi [\aap] {10.1051/0004-6361/201731982},
  \href {http://adsabs.harvard.edu/abs/2018A%26A...613A..46D} {613, A46}

\bibitem[\protect\citeauthoryear{{Dolez} \& {Vauclair}}{{Dolez} \&
  {Vauclair}}{1981}]{1981A&A...102..375D}
{Dolez} N.,  {Vauclair} G.,  1981, \aap, \href
  {http://adsabs.harvard.edu/abs/1981A%26A...102..375D} {102, 375}

\bibitem[\protect\citeauthoryear{{Fontaine} \& {Brassard}}{{Fontaine} \&
  {Brassard}}{2008}]{2008PASP..120.1043F}
{Fontaine} G.,  {Brassard} P.,  2008, \mn@doi [\pasp] {10.1086/592788}, \href
  {http://adsabs.harvard.edu/abs/2008PASP..120.1043F} {120, 1043}

\bibitem[\protect\citeauthoryear{{Giammichele}, {Fontaine}, {Brassard}  \&
  {Charpinet}}{{Giammichele} et~al.}{2014}]{2014IAUS..301..285G}
{Giammichele} N.,  {Fontaine} G.,  {Brassard} P.,   {Charpinet} S.,  2014, in
  {Guzik} J.~A.,  {Chaplin} W.~J.,  {Handler} G.,   {Pigulski} A.,  eds,  IAU
  Symposium Vol. 301, Precision Asteroseismology. pp 285--288,
  \mn@doi{10.1017/S1743921313014464}

\bibitem[\protect\citeauthoryear{{Giammichele}, {Fontaine}, {Brassard}  \&
  {Charpinet}}{{Giammichele} et~al.}{2016}]{2016ApJS..223...10G}
{Giammichele} N.,  {Fontaine} G.,  {Brassard} P.,   {Charpinet} S.,  2016,
  \mn@doi [\apjs] {10.3847/0067-0049/223/1/10}, \href
  {http://adsabs.harvard.edu/abs/2016ApJS..223...10G} {223, 10}

\bibitem[\protect\citeauthoryear{{Goldreich} \& {Wu}}{{Goldreich} \&
  {Wu}}{1999}]{1999ApJ...511..904G}
{Goldreich} P.,  {Wu} Y.,  1999, \mn@doi [\apj] {10.1086/306705}, \href
  {http://adsabs.harvard.edu/abs/1999ApJ...511..904G} {511, 904}

\bibitem[\protect\citeauthoryear{{Herwig}}{{Herwig}}{2000}]{2000A&A...360..952H}
{Herwig} F.,  2000, \aap, \href
  {http://adsabs.harvard.edu/abs/2000A%26A...360..952H} {360, 952}

\bibitem[\protect\citeauthoryear{{Karakas} \& {Lattanzio}}{{Karakas} \&
  {Lattanzio}}{2014}]{2014PASA...31...30K}
{Karakas} A.~I.,  {Lattanzio} J.~C.,  2014, \mn@doi [pasa]
  {10.1017/pasa.2014.21}, \href
  {http://adsabs.harvard.edu/abs/2014PASA...31...30K} {31, e030}

\bibitem[\protect\citeauthoryear{{Kunz}, {Fey}, {Jaeger}, {Mayer}, {Hammer},
  {Staudt}, {Harissopulos}  \& {Paradellis}}{{Kunz}
  et~al.}{2002}]{2002ApJ...567..643K}
{Kunz} R.,  {Fey} M.,  {Jaeger} M.,  {Mayer} A.,  {Hammer} J.~W.,  {Staudt} G.,
   {Harissopulos} S.,   {Paradellis} T.,  2002, \mn@doi [\apj]
  {10.1086/338384}, \href {http://adsabs.harvard.edu/abs/2002ApJ...567..643K}
  {567, 643}

\bibitem[\protect\citeauthoryear{{Romero}, {C{\'o}rsico}, {Althaus}, {Kepler},
  {Castanheira}  \& {Miller Bertolami}}{{Romero}
  et~al.}{2012}]{2012MNRAS.420.1462R}
{Romero} A.~D.,  {C{\'o}rsico} A.~H.,  {Althaus} L.~G.,  {Kepler} S.~O.,
  {Castanheira} B.~G.,   {Miller Bertolami} M.~M.,  2012, \mn@doi [\mnras]
  {10.1111/j.1365-2966.2011.20134.x}, \href
  {http://adsabs.harvard.edu/abs/2012MNRAS.420.1462R} {420, 1462}

\bibitem[\protect\citeauthoryear{{Romero}, {Kepler}, {C{\'o}rsico}, {Althaus}
  \& {Fraga}}{{Romero} et~al.}{2013}]{2013ApJ...779...58R}
{Romero} A.~D.,  {Kepler} S.~O.,  {C{\'o}rsico} A.~H.,  {Althaus} L.~G.,
  {Fraga} L.,  2013, \mn@doi [\apj] {10.1088/0004-637X/779/1/58}, \href
  {http://adsabs.harvard.edu/abs/2013ApJ...779...58R} {779, 58}

\bibitem[\protect\citeauthoryear{{Saio}}{{Saio}}{2013}]{2013EPJWC..4305005S}
{Saio} H.,  2013, in European Physical Journal Web of Conferences. p. 05005,
  \mn@doi{10.1051/epjconf/20134305005}

\bibitem[\protect\citeauthoryear{{Tremblay}, {Bergeron}  \&
  {Gianninas}}{{Tremblay} et~al.}{2011}]{2011ApJ...730..128T}
{Tremblay} P.-E.,  {Bergeron} P.,   {Gianninas} A.,  2011, \mn@doi [\apj]
  {10.1088/0004-637X/730/2/128}, \href
  {http://adsabs.harvard.edu/abs/2011ApJ...730..128T} {730, 128}

\bibitem[\protect\citeauthoryear{{Winget} \& {Kepler}}{{Winget} \&
  {Kepler}}{2008}]{2008ARA&A..46..157W}
{Winget} D.~E.,  {Kepler} S.~O.,  2008, \mn@doi [\araa]
  {10.1146/annurev.astro.46.060407.145250}, \href
  {http://adsabs.harvard.edu/abs/2008ARA%26A..46..157W} {46, 157}

\bibitem[\protect\citeauthoryear{{Winget}, {van Horn}, {Tassoul}, {Fontaine},
  {Hansen}  \& {Carroll}}{{Winget} et~al.}{1982}]{1982ApJ...252L..65W}
{Winget} D.~E.,  {van Horn} H.~M.,  {Tassoul} M.,  {Fontaine} G.,  {Hansen}
  C.~J.,   {Carroll} B.~W.,  1982, \mn@doi [\apjl] {10.1086/183721}, \href
  {http://adsabs.harvard.edu/abs/1982ApJ...252L..65W} {252, L65}

\makeatother
\end{thebibliography}
